\documentclass[aps,prl,floatfix,superscriptaddress,reprint]{revtex4-2}
\usepackage{mathrsfs,braket}
\usepackage{amssymb, amsbsy, amsmath, latexsym, dsfont, array, layout, graphicx, mathrsfs, color, ulem, bm}
\usepackage[colorlinks,linkcolor=blue,anchorcolor=red,citecolor=blue]{hyperref}

\begin{document}
\title{Family of  one-dimensional self-dual quasicrystals with critical phases}

 \author{Wenzhi Wang}
 \affiliation{Laboratory of Quantum Information, University of Science and Technology of China, Hefei 230026, China}
\author{Wei Yi}
\email{wyiz@ustc.edu.cn}
\affiliation{Laboratory of Quantum Information, University of Science and Technology of China, Hefei 230026, China}
\affiliation{Anhui Province Key Laboratory of Quantum Network, University of Science and Technology of China, Hefei 230026, China}
\affiliation{CAS Center For Excellence in Quantum Information and Quantum Physics, Hefei 230026, China}
\affiliation{Hefei National Laboratory, University of Science and Technology of China, Hefei 230088, China}
\affiliation{Anhui Center for Fundamental Sciences in Theoretical Physics, University of Science and Technology of China, Hefei 230026, China}
\author{Tianyu Li}
\email{tianyuli@m.scnu.edu.cn}
\affiliation {Key Laboratory of Atomic and Subatomic Structure and Quantum Control (Ministry of Education), Guangdong Basic Research Center of Excellence for Structure and Fundamental Interactions of Matter, School of Physics, South China Normal University, Guangzhou 510006, China}
\affiliation {Guangdong Provincial Key Laboratory of Quantum Engineering and Quantum Materials, Guangdong-Hong Kong Joint Laboratory of Quantum Matter, Frontier Research Institute for Physics, South China Normal University, Guangzhou  510006, China}

\begin{abstract}
We propose a general framework for constructing self-dual one-dimensional quasiperiodic lattice models with arbitrary-range hoppings and multifractal behaviors.
Our framework generates a broad spectrum of one-dimensional quasicrystals, ranging from the off-diagonal Aubry-Andr\'{e}-Harper models on one end, to those featuring long-range hoppings with varied quasiperiodic modulations on another.
Focusing on models with off-diagonal quasiperiodic hoppings with power-law decay,
we exploit the fact that, when the self-dual condition is satisfied, the system must be in the critical state with multifractal properties.
This enables the engineering of models with competing extended, critical, and localized phases, with richly structured mobility edges separating them.
As an outstanding example, we show that a limiting case of our family of self-dual quasicrystals can be implemented using Rydberg-atom arrays.
Our work offers a systematic route toward critical phases from self-duality considerations, and would facilitate the experimental simulation of these exotic states.
\end{abstract}

\maketitle

{\it Introduction.---}
Anderson localization is a fundamental quantum phenomenon in which disorder leads to the exponential localization of eigenstates, thereby inhibiting wave diffusion~\cite{anderson,Lee1985,wiersma,Kramer1993,Evers2008}.
The associated location-delocalization transition generally depends on the spatial dimensions,
for instance, an infinitesimally weak disorder is sufficient to induce Anderson localization in one dimension~\cite{DJThouless1974,Four,Hetenyi}.
But Anderson localization also arises in quasiperiodic models, where, in a one dimensional quasicrystal for instance, the localization only occurs beyond a critical threshold of the quasiperiodic potential.
This is exemplified by the well-known Aubry-Andr\'{e}-Harper (AAH) model~\cite{AA}, characterized by the eigen equation
$t\left(\psi_{n+1}+\psi_{n-1}\right)+V \cos (2 \pi \tau n) \psi_n=E \psi_n$,
where $\psi_n$ is the wave function component on site $n$, $t$ is the nearest-neighbor hopping strength, $V$ is the amplitude of the quasiperiodic potential, $\tau$ is an irrational number, and $E$ is the eigenenergy.
Crucially, the AAH model exhibits self-duality, such that it Fourier-transforms onto itself but with switched parameters $V\leftrightarrow 2t$. The self-dual point $V=2t$ thus marks a sharp transition between the extended (for $V<2t$) and localized (for $V>2t$) phases.
Remarkably, at the self-dual point, the system becomes critically localized~\cite{critical1,critical2,critical3,critical4,critical5,critical6,santos,lizhi,LEP,gopa,zhouxinchi}.

Further engineering of the AAH model can have two important consequences. First, with the addition
of quasiperiodic hopping
coefficients, the critical point can be expanded into a critical phase occupying a broad region on the phase diagram.
These critical states exhibit unique spectral statistics~\cite{spectral1,spectral2,spectral3}, multifractal properties~\cite{multi1,multi2,multi3},  unconventional dynamics~\cite{dynamical1,dynamical2,dynamical3}, and superconductivity~\cite{supercon1,supercon2}. Recent studies have shown that when interactions are introduced into
systems supporting single-particle critical states, the resulting models
can host many-body critical phases, representing a dynamical regime that
is neither ergodic nor many-body localized. ~\cite{wangyucheng,lixiao,roy}.
Second, by introducing long-range hopping terms~\cite{dassarma1} or fine-tuned on-site quasiperiodic potentials~\cite{dassarma2}, mobility edges emerge in the eigenspectum, where critical energies separate extended and localized states~\cite{wiersma,Evers2008,dassarma1,dassarma2,dassarma3,lixiao,xiongjun,gongming}.
Tunable mobility edges suggest energy- or
parameter-dependent transport, which are highly desirable for device design.
These possibilities have thus sparked widespread interest, including theoretical studies on periodically (or quasiperiodically) driven systems~\cite{qua1,qua2,qua3,qua4,qua5,qua6} and non-Hermitian quasicrystals~\cite{non1,non2,non3,non4,non5,non6,non7,non8,non9,gaoxianlong,nonoffaah}, as well as experimental implementations using ultracold atoms~\cite{cold1,cold2,cold3,cold4,cold5,cold6,cold7,cold8,cold9,cold10,cold11}, photonic crystals~\cite{latt1,latt2,latt3,latt4,latt5,latt6,latt7}, optical cavities~\cite{cavity1,cavity2}, and superconducting circuits~\cite{super1,critical1}.
Despite these activities, the general mechanisms responsible for the emergence of critical states remain an open question, and the coexistence (in the same phase diagram) of extended, critical, and localized phases has so far been observed only in a handful of highly fine-tuned models~\cite{dual,flux,coexist,lizhi}.
While it is tempting to address these issues from the perspective of self-duality, only a limited number of quasiperiodic models are known to be self-dual~\cite{gopa,dassarma1,dassarma2,dual}.

In this work, we propose a general framework for constructing self-dual one-dimensional quasiperiodic models with with arbitrary hopping modulations that give rise to a variety of critical behaviors.
Similar to the AAH example, with our scheme, a model constructed with the parameters $(a,b)$ is dual to the same model but parametrized by $(b,a)$. This naturally leads to the self-dual condition $a=b$, where the system must be in the critical state, and remains so in its vicinity.
The family of quasicrystals so generated spans a wide spectrum, with the (generalized) AAH models on one end, and those with long-range hoppings having exponential or power-law decay
on another.
We focus on models exhibiting off-diagonal quasiperiodic hoppings with power-law decay,
and show that the overall phase diagram is symmetric with respect to the self-dual condition $a=b$, featuring rich mobility edges in between the extended, critical, and localized phases.
We then propose an experimental-simulation scheme using Rydberg-atom arrays.
Our work provides a general scheme for engineering self-dual quasicrystals, as well as a route toward critical phases from self-duality considerations.

{\it Self-dual quasicrystals.---}
We consider a family of one-dimensional tight-binding models
\begin{align}
 \hat{H}(a,b)=\sum_{m\neq n}\hat{b}_{m}^{\dagger}\hat{b}_{n}f_{|m-n|}(a)F_{m+n}(b),\label{eq:Habmain}
\end{align}
where $\hat{b}_n^\dagger$ ($\hat{b}_n$) are the creation (annihilation) operators on site $n$. The hopping coefficients are determined by the product of two functions, $f_{|m-n|}(a)$ and $F_{m+n}(b)$, which depend on the site index and the real parameters $a$ and $b$. Furthermore, we require that the two functions be related according to
\begin{align}
F_x(a)=\sum_{s=1}^{+\infty}f_{s}(a)\cos(\tau s\pi x), \label{eq:Fourier}
\end{align}
where the irrational number $\tau=({\sqrt{5}-1})/{2}$. Note that our proposal is still applicable when choosing a different irrational number.
Performing the dual transformation
\begin{equation}
\tilde{\hat{b}}_{k}=\frac{1}{\sqrt{N}}\sum_{m=1}^{N}e^{-2i\pi\tau mk}\hat{b}_{m},\label{eq:trans_b}
\end{equation}
where $N$ represents the total number of sites, the Hamiltonian in the dual space reads
\begin{equation}
 \hat{H}(a,b)=\sum_{k\neq l}\tilde{\hat{b}}_{k}^{\dagger}\tilde{\hat{b}}_{l}f_{|k-l|}(b)F_{k+l}(a).\label{eq:H_d}
\end{equation}
Here the derivation of Hamiltonian~(\ref{eq:H_d}) only holds for an irrational $\tau$~\cite{supp}.
Equations~(\ref{eq:Habmain}) and (\ref{eq:H_d}) establish that $\hat{H}(a,b)$ and $\hat{H}(b,a)$ are dual to each other, such that $a=b$ defines the self-dual condition.
It follows that, if the $j$th eigenstate of $\hat{H}(a,b)$, denoted as $\psi^{(j)}$, is localized, then the $j$th eigenstate of $\hat{H}(b,a)$, denoted as $\tilde\psi^{(j)}$, must be extended, and vice versa.
Furthermore, if $\psi^{(j)}$ is critical, then $\tilde\psi^{(j)}$ is also critical, such that all eigenstates of $\hat{H}(a,a)$ are critical, being at the self-dual point $a=b$. Note that, in the following, we index the eigenstates according to their eigenenergies in ascending order.

Importantly, Hamiltonian $\hat{H}(a,b)$ is determined by the functions $\{f_s(a)\}_{s=1,2,...,+\infty}$, which are arbitrary, and subject only to the constraint that the summation in Eq.~(\ref{eq:Fourier}) converge.
Such a flexibility allows us to construct a broad class of self-dual quasicrystals with multifractal properties~\cite{supp}.

As a notable special case, we set $f_1(a) = a$, and $f_s(a)=0$ for $s\neq 1$. We then have
\begin{align}
F_x(a)=a \cos(\tau\pi x),\label{eq:F_off_AAH}
\end{align}
which leads to the Hamiltonian
\begin{equation}
 \hat{H}(a,b)=\sum_{n}ab \cos[\tau\pi(2n+1)](\hat{b}_{n+1}^{\dagger}\hat{b}_{n}+\hat{b}_{n}^{\dagger}\hat{b}_{n+1}).\label{eq:H_off_AAH}
\end{equation}
This is the well-studied off-diagonal AAH model~\cite{nonoffaah,offaah}. Since Eq.~(\ref{eq:H_off_AAH}) remains invariant under the dual transformation, with $\hat{H}(a,b)=\hat{H}(b,a)$, all eigenstates of the off-diagonal AAH model are critical. Note that the conventional AAH model can also be generated based on our construction~\cite{supp}.

{\it Quasicrystals with power-law hoppings.---}
Inspired by the recent progress in Rydberg-atom arrays with long-range interactions, we now focus on
a class of quasiperiodic models with power-law hopping terms.
Specifically, we consider $f_s(a) = \frac{1}{s^a}$ for $s\leq d$, and $f_s(a)=0$ for $s> d$. In this case, we have
\begin{align}
F_{x}(a)=\sum_{s=1}^{d}\frac{1}{s^{a}}\cos(\tau s\pi x),\label{eq:F_power}
\end{align}
and the corresponding Hamiltonian is
\begin{align}
\hat{H}_p(a,b)=\sum_{1\leq|m-n|\leq d}\hat{b}_{m}^{\dagger}\hat{b}_{n}\frac{1}{|m-n|^{a}}\sum_{s=1}^{d}\frac{1}{s^{b}}\cos[\tau s\pi (m+n)],\label{eq:Hsd_power}
\end{align}
where $d$ denotes the maximum hopping range, while $a$ and $b$ respectively characterize the exponent of the power-law decay in hopping, and the amplitude modulation. Note that for $a,b\to \infty$, Hamiltonian (\ref{eq:Hsd_power}) reduces to the off-diagonal AAH model (\ref{eq:H_off_AAH}).

To corroborate the self-duality and localization properties of the model, we study the fractal dimension ($\text{FD}$) of $\hat{H}_p(a,b)$. The FD is given by
\begin{equation}
\text{FD}=-\frac{\ln(\text{IPR}_j)}{\ln(N)},\label{eq:FD}
\end{equation}
where the inverse participation ratio of the $j$th eigenstate is defined as
$\text{IPR}_j=\sum_{n}|\psi^{(j)}_{n}|^{4},$ with $\psi^{(j)}_n$ indicating the component on site $n$ of the $j$th eigenstate.
In the thermodynamic limit, $\text{FD}$ approaches 1 for extended states, and 0 for localized states, while $0<\text{FD}<1$ signifies the critical states~\cite{hoffmann,zhouxinchi}.

\begin{figure}[tbp]
\centering
\includegraphics[scale=0.45]{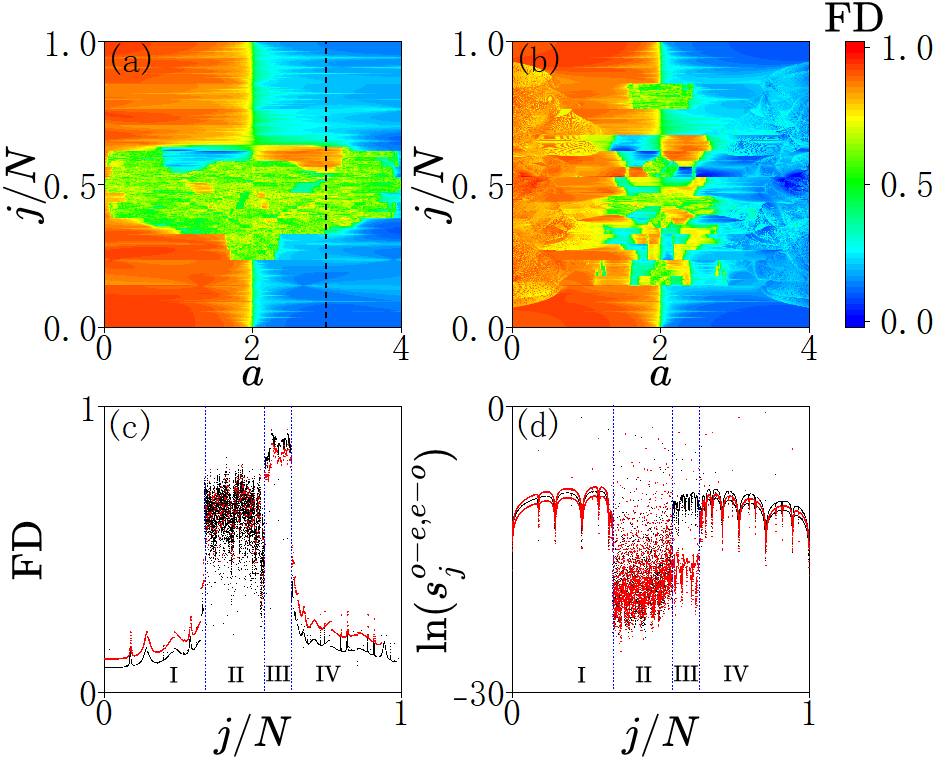}
\caption{Fractal dimension $\text{FD}$ of $\hat{H}_p(a,4-a)$ for (a) $d=2$, and (b) $d=10$.
(c) The FD as a function of the eigenstate index
$j/N$, with fixed $a=3$ [the vertical black dashed line in (a)], for
$N= 2584$ (red dots) and $N= 46\,368$ (black dots), respectively.
(d) The even-odd eigen-level spacings $s_{j}^{e-o}$ for even $j$ (black dots) and $s_{j}^{o-e}$ for odd $j$ (red dots), both as functions of $j/N$ for $N= 28\,657$.
The vertical blue dashed lines in (c) and (d) indicate the positions of the mobility edge, separating four distinct regions corresponding to the localized (I, IV), critical (II), and extended (III) phases.
We take $N=2584$ in (a)(b), and impose the periodic boundary condition for all calculations.
For numerical calculations, we approximate the golden ratio $\tau$ as the ratio of two adjacent $u$-Fibonacci numbers, with $\tau=1597/2584$ here~\cite{gaoxianlong,supp}.
\label{fig:1}}
\end{figure}

As a convenient sampling of the parameter space, we impose the constraint $a+b=4$, such that the self-dual condition is satisfied for $a=2$.
In Fig.~\ref{fig:1}(a)(b), we show the computed FD as functions of $a$ and the eigenstate index $j/N$ for different hopping ranges $d$.
The resulting phase diagram exhibits a rich structure, featuring all three phases: the extended (red), the critical (green), and the localized (blue).
Consistent with the self-dual properties of the model, the phase diagram is symmetric:
regions of the extended (localized) phase to the left of $a=2$ are symmetric to those of the localized (extended) phase to the right, while regions of the critical phase are symmetric with respect to $a=2$.
With increasing hopping range $d$, the stability region of the critical phase shrinks and becomes fragmented.
Note that hoppings with power-law decay are not essential for our scheme,
other forms of decay (exponential for instance) can lead to qualitatively similar phase diagrams~\cite{supp}.

In Fig.~\ref{fig:1}(c), we show the FD as a function of the eigenstate index $j/N$ for different system sizes. In the localized regions (marked I and IV), $\text{FD}$ tends toward zero with increasing system size, while in the extended region (marked III), $\text{FD}$ increases toward unity. In contrast, in the critical region (marked II), $\text{FD}$ remains between 0 and 1, showing minimal dependence on the system size.

Alternatively, the three phases can be distinguished by analyzing the level statistics.
Denoting the ordered eigenenergies as $\{ E_j\}_{j=1,2,...N}$, we define the even-odd eigen-level spacing as  $s_{j}^{e-o}= E_{j}-E_{j-1}$ for even $j$, and $s_{j}^{o-e}=E_{j}- E_{j-1}$ for odd $j$~\cite{qua5,santos}. In the extended phase, the spectrum is nearly doubly degenerate, leading to $s_{j}^{o-e}\sim 0$, so that a noticeable gap exists between $s_{j}^{e-o}$ and $s_{j}^{o-e}$. In the localized phase, such degeneracies vanish, and the numerically calculated gap sharply decreases in magnitude. In the critical phase, both $s_{j}^{e-o}$ and  $s_{j}^{o-e}$ exhibit strongly scattered distributions, indicative of its multifractal properties~\cite{santos}.
These expected features are clearly visible in Fig.\ref{fig:1}(d), with the phase boundaries consistent with those in Figs.~\ref{fig:1}(a) and \ref{fig:1}(c).

\begin{figure}[tbp]
\centering
\includegraphics[scale=0.45]{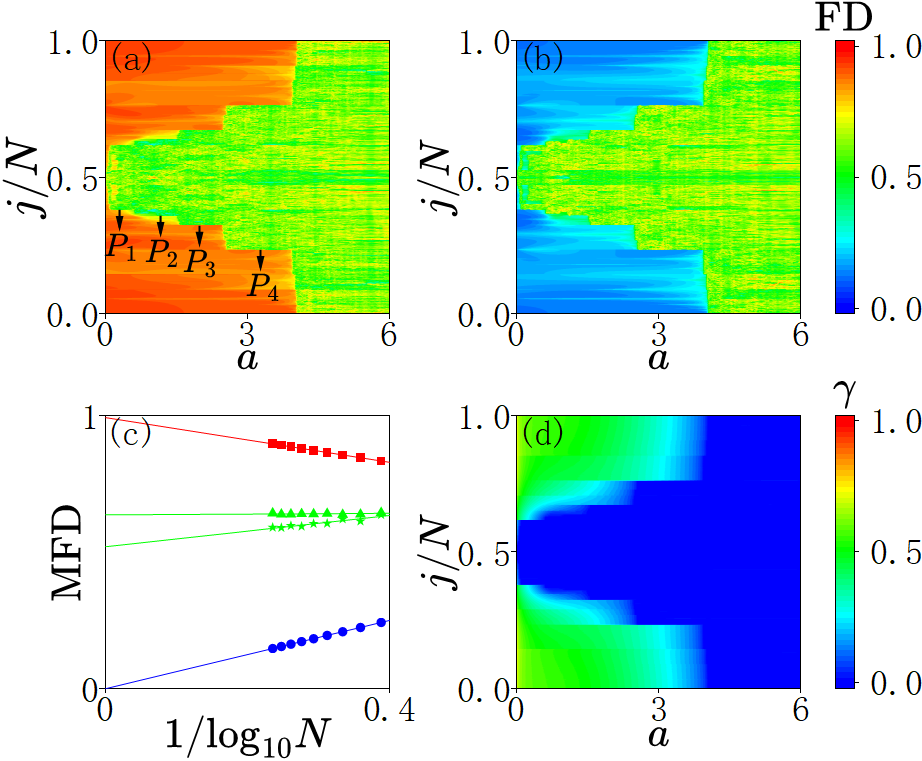}
\caption{
The FDs for different eigenstates with $d=2$ for (a) $\hat{H}_p(a,+\infty)$ and (b) $\hat{H}_p(+\infty,a)$.
Four characteristic regions with step-wise mobility edges in $j/N$ can be identified, labeled by $j/N=P_{1,2,3,4}$.
(c) MFDs as functions of $1/\log_{10}N$ for eigenstates in different spectral regions for $a=3$ in (a)(b). Red squares and blue circles correspond to the extended [in (a)] and localized phases [in (b)], respectively. Green triangles and stars indicate the MFDs for the critical states in (a) and (b), respectively. Solid lines denote results of linear fit.
(d) The Lyapunov exponent as a function of $a$ and $j/N$. We take $N=2584$ in (a)(b)(d), and impose the periodic boundary condition for all calculations. \label{fig:2}}
\end{figure}

\begin{figure}[tbp]
\centering
\includegraphics[scale=0.5]{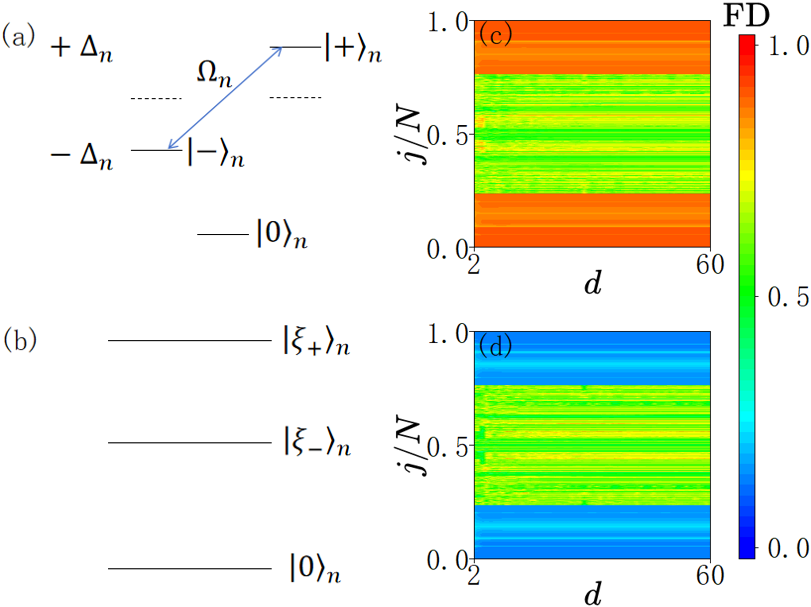}
\caption{
(a) Level scheme of the microwave-dressed Rydberg states of the $n$th atom in the one-dimensional array. Here $2\Delta_n$ is the site-dependent energy splitting between the Rydberg states $|+\rangle_n$ and $|-\rangle_n$, $\Omega_n$ is the Rabi frequency of the coupling between $|+\rangle_n$ and $|-\rangle_n$.
(b) Level scheme for the microwave-dressed basis of the $n$th atom.
(c)(d) Fractal dimension $\text{FD}$ of different eigenstates with varying interaction (hopping) range $d$ for (c) $H(3,+\infty)$ and (d) $H(+\infty,3)$, respectively.
The two mobility edges are located at $j/N = 2-2\tau$ and $2\tau-1$, respectively.
For all calculations, the system size is $N=2584$ and  we impose the periodic boundary condition.
\label{fig:3}}
\end{figure}

Next, we consider a particular limiting case of Eq.~(\ref{eq:Hsd_power})
\begin{equation}
 \hat{H}_p(a,+\infty)=\sum_{1\leq|m-n|\leq d}\hat{b}_{m}^{\dagger}\hat{b}_{n}\frac{\cos[\tau\pi(m+n)]}{|m-n|^{a}},\label{eq:Hsd_power_a_inf}
\end{equation}
whose dual is given by
\begin{equation}
\hat{H}_p(+\infty,a)=\sum_{m}(\hat{b}_{m}^{\dagger}\hat{b}_{m+1}+\hat{b}_{m+1}^{\dagger}\hat{b}_{m})\sum_{s=1}^{d}\frac{\cos[\tau s\pi (2m+1)]}{s^{a}}.\label{eq:Hsd_power_inf_a}
\end{equation}
Note that $\hat{H}(a,+\infty)$ features long-range hoppings with a single quasiperiodicity, whereas its dual $\hat{H}(+\infty,a)$ involves only nearest-neighbor hopping, modulated by a sum of quasiperiodic components.

In Figs.~\ref{fig:2}(a) and \ref{fig:2}(b), we show the FDs of the eigenstates of
$\hat{H}(a,+\infty)$ and $\hat{H}(+\infty,a)$, respectively.
While $\hat{H}(a,+\infty)$  possesses only extended and critical phases, its dual $\hat{H}(+\infty,a)$ supports localized and critical phases.
In either case, the phase diagram is symmetric with respect to $j/N = 1/2$, which originates from a spectral symmetry: for each eigenvalue $E$, its opposite eigenvalue $-E$ also exists in the spectrum, with both eigenstates featuring the same spatial localization since they differ only by a sign reversal on odd (or even) lattice sites~\cite{supp}.
Furthermore, regions with mobility edges that are constant in $j/N=P_{1,2,3,4}$ can be identified, with $P_1=1-\tau$, $P_2=20\tau-12$, $P_3=7\tau -4$, and $P_4=2\tau-1$, all independent of the system size.  Here the expressions of $P_n$ are determined through numerical analysis, with their forms sensitive to the choice of $\tau$~\cite{supp}.
Further, although the positions of these mobility edges remain unchanged in terms of the eigenstate index $j/N$ within each region in Fig.~\ref{fig:2}, their corresponding eigenenergies vary with $a$.
These mobility edges separate the extended (red) or localized (blue) phases from the critical phase (green).
For a fermionic system, the step-wise mobility edges in $j/N$ suggest a density-dependent quantum transport that can be robustly tuned through the parameter $a$.

To better confirm the nature of different phases, we perform a finite-size scaling of the mean fractal dimension (MFD). Take $a=3$ as an example, the MFD in the critical region is defined by averaging the FD of all eigenstates with $j/N \in [P_4, 1 - P_4]$. The
MFD in the extended (localized) region is averaged outside this interval.
Figure~\ref{fig:2}(c) shows the MFDs as functions of $1/\log_{10}N$ for different phases of
$\hat{H}_p(3,+\infty)$ and $\hat{H}_p(+\infty,3)$, respectively.
Crucially, in the thermodynamic limit $1/\log_{10}N \to 0$, the MFDs in the critical regions extrapolate to finite values in $(0,1)$, consistent with the multifractal nature of the critical states and confirming the critical nature of the model away from the self-dual point.

Lyapunov exponents (LEs) offer another useful diagnostic for localization.
For an eigenstate with energy $E$, the LE is defined as~\cite{lizhi,LEP}
\begin{equation}
\gamma(E)=\lim_{N\rightarrow\infty}\frac{1}{N}\text{ln}||\Pi_{n=1}^{N}T_{n}||,\label{eq:LE}
\end{equation}
where $||\cdot||$ denotes the matrix norm, corresponding to taking the largest absolute value among the eigenvalues. The transfer matrix
$T_{n}$ is given by
\begin{align}
T_{n}=\begin{pmatrix}\frac{E}{t_{n}} & -\frac{t_{n-1}}{t_{n}}\\
1 & 0
\end{pmatrix},
\end{align}
with $t_n = \sum_{s=1}^{d}\frac{\cos[\tau s\pi (2n+1)]}{s^{a}}$.
Here the calculation is feasible since the transfer matrices are easily accessible for $\hat{H}_p(a,+\infty)$ and $\hat{H}_p(+\infty,a)$, which is not generally the case for $\hat{H}_p(a,b)$.
In Fig.~\ref{fig:2}(d), we show the calculated LE as a function of $a$ and $j/N$, the results match excellently with the phase diagram in Fig.~\ref{fig:2}(b). Specifically, for critical states, the LE vanishes ($\gamma=0$), while for localized states, it remains finite~\cite{LEP}.

{\it Experimental realization using Rydberg atoms.---}
The dipole-dipole interactions in Rydberg atoms dictate a $1/r^3$ long-range power-law decay. Based on such an observation, we illustrate below that the model in
Eq.~(\ref{eq:Hsd_power_a_inf}) can be experimentally realized in a Rydberg-atom array for $a=3$.

Consider a one-dimensional array of $^{87}$Rb atoms for instance. As illustrated in Fig.~\ref{fig:3}(a), the internal-state space of the atom on the $n$th site consists of the Rydberg states $|60S_{1/2},m_j=1/2\rangle$, denoted as $|0\rangle_n$, as well as $|60P_{3/2},m_j=3/2\rangle$ and $|60P_{3/2},m_j=-1/2\rangle$, denoted as $|+\rangle_n$ and $|-\rangle_n$, respectively.
A site-dependent energy splitting of $2\Delta_n$ is imposed between the states $|+\rangle_n$ and $|-\rangle_n$, which can be induced by a magnetic-field gradient.
These two states are further coupled by a site-dependent microwave field with Rabi frequency $\Omega_n$. By moving to the eigenbasis of the microwave-dressed states~\cite{supp}, the internal degrees of freedom of the atom on the $n$th site are labeled $\{|0\rangle_n, |\xi_\pm\rangle_n\}$, as shown in Fig.~\ref{fig:3}(b).

Performing the rotating-wave approximation in an appropriate rotating frame, and setting appropriate
values of $\Delta_n$ and $\Omega_n$, we reproduce Hamiltonian~(\ref{eq:Hsd_power_a_inf}) as~\cite{Sigma,supp}
\begin{equation}\
\hat{H}_{\xi_{\pm}}=A\sum_{m\neq n}\frac{\cos[\pi\tau(m+n)]}{|m-n|^{3}}\hat{b}_{\xi_{+},m}^{\dagger}\hat{b}_{\xi_{+},n},\label{eq:H_GAAHmain}
\end{equation}
where $\hat{b}_{\xi_{+},n}^{\dagger}=|\xi_{+}\rangle_{n}\langle0|_{n}$. For the derivation, we require that the energy splittings between the microwave-dressed basis be much larger than the dipole interaction energy~\cite{supp}, and focus on the case where only a single excited state $|\xi_{+}\rangle$ is present. Here the pre-factor $A$ is proportional to the dipolar interaction energy between adjacent atoms.
Under typical experimental conditions, it suffices to take the hopping range $d=2$ or $3$~\cite{experid1,experid2,experid3}.

In Figs.~\ref{fig:3}(c) and \ref{fig:3}(d), we show the FD of different eigenstates of the Hamiltonian (\ref{eq:H_GAAHmain}) and its dual model, with varying interaction range $d$.
Notably, for Eq.~(\ref{eq:H_GAAHmain}) [see Fig.~\ref{fig:3}(c)], two mobility edges separate the extended and critical phases, and they remain unchanged in terms of the eigenstate index $j/N$ across all values of $d$. For its dual model, two mobility edges similarly separate the localized and critical phases, which also remain unchanged in terms of $j/N$ for all values of $d$ [see Fig.~\ref{fig:3}(d)].
The flat mobility edges in $d$ derive from the fast $1/r^3$ decay of the hopping terms in Eq.~(\ref{eq:H_GAAHmain}). This is in contrast to Fig.~\ref{fig:1}, where the constraint $a+b=4$ renders either the hopping or its amplitude modulation decay slower in $d$.
Experimentally, the different phases can be identified through local dynamic observables~\cite{supp,super1}.

{\it Discussions.---}
Our work offers a systematic route toward self-duality, and, by linking self-duality with critical states, provides a practical access to the latter.
The practice is a valuable complement to recent studies of critical states~\cite{dual,flux,lizhi}, where analytical insights are often obtained for highly specialized models.
While a limiting case of our scheme can be readily implemented with Rydberg atoms,
the highly tunable interaction and lattice parameters of the Rydberg-atom arrays open up a plethora of possibilities for the simulation of critical states and their many-body counterparts.
The latter is possible by considering more than one excitations in the Rydberg-atom array.
For future studies, it would be interesting to search for analytically solvable models within our framework, and extend the scheme to higher dimensions.

\acknowledgments
This work is supported by the National Natural Science Foundation of China (Grant No. 12374479), and the Innovation Program for Quantum Science and Technology (Grant No. 2021ZD0301904).

\end{document}